\documentclass[aps,prb,twocolumn,groupedaddress,showpacs]{revtex4}
\usepackage{graphicx}
\usepackage{amssymb}
\begin{document}

\title{Influence of the sample geometry on the vortex matter in superconducting microstructures}

\author{M. Morelle}
\email[]{mathieu.morelle@fys.kuleuven.ac.be} \homepage[]{http://www.fys.kuleuven.ac.be/vsm/nsm} \affiliation{Nanoscale Superconductivity and Magnetism
Group, Laboratory for Solid State Physics and Magnetism, K.U. Leuven, Celestijnenlaan 200D, B-3001 Leuven, Belgium}
\author{J. Bekaert}
\altaffiliation{Currently at IMEC, Kapeldreef 75, B-3001 Leuven, Belgium}
\author{V. V. Moshchalkov}
\affiliation{Nanoscale Superconductivity and Magnetism Group, Laboratory for Solid State Physics and Magnetism, K.U. Leuven, Celestijnenlaan 200D, B-3001
Leuven, Belgium}
\date{\today}

\begin{abstract}
The dependence of the vortex penetration and expulsion on the geometry of mesoscopic superconductors is reported. Hall magnetometry measurements were
performed on a superconducting Al square and triangle. The stability of the vortex patterns imposed by the sample geometry is discussed. The
field-temperature $H-T$ diagram has been reconstructed showing the transitions between states with different vorticity. We have found that the vortex
penetration is only weakly affected by the vortex configuration inside the sample while the expulsion is strongly controlled by the stability of the
vortex patterns. A qualitative explanation for this observation is given.
\end{abstract}

\pacs{74.78.Na, 74.25.Dw, 74.25.Ha}

\maketitle
\section{Introduction}

In samples with the size comparable to the coherence length, the confinement of the superconducting condensate determines the arrangement of the
vortices. Recently, Chibotaru~{\it et al.\/}\cite{chibotaru00,chibotaru01} have shown that new vortex patterns can arise to preserve the symmetry of the
sample. This discovery has stimulated the research on the appearance of additional vortex-antivortex pairs and their stability in squares, triangles and
rectangles\cite{bonca01,melnikov02,misko02PhysC,misko02,teniers03,mertelj03}. While the magnetization of a superconducting disk has already been studied
experimentally~\cite{geim97nat1}, only transport measurements have been performed so far to characterize the phase boundary of different mesoscopic
polygons~\cite{vvm95nature,morelle02,morelle03}. The presence of sharp corners in these samples enhances locally the superconducting order
parameter\cite{fomin98epl,fomin99err,schweigert99prbrief}.

The recent process in nanopatterning has made it possible to fabricate very sensitive Hall sensors made from the GaAs/AlGaAs semiconductor
heterostructure containing a two-dimensional electron gas. These Hall sensors and Hall magnetometers have been efficiently used for characterization of
both ferromagnetic~\cite{schuh01,bekaert02} and superconducting~\cite{geim97apl} materials as well as for local visualization of the magnetic field
profile\cite{bekaert02} in a non-invasive way. A submicron superconducting or ferromagnetic element can be deposited on the sensitive area of the probe.
The magnetic induction is measured from the Hall voltage. It has been calculated\cite{peeters98} that the local induction can be considered as the
averaged induction over the total sensing area of the probe.

Using Hall magnetometers to measure the magnetization of superconducting mesoscopic disks,  a paramagnetic signal was observed\cite{geim98nat}. The Hall
magnetometers have been studied theoretically in the ballistic\cite{peeters98} and in the diffusive\cite{ibrahim98} regime.  It was shown that the size
of the detector has a substantial influence on the magnitude of the measured magnetization and can even change the sign of it\cite{deo99prb}.

Hall magnetometry was also successfully used to investigate the paramagnetic Meissner effect. This effect has been studied in high temperature
superconductors. It was also observed in conventional low temperature superconductors\cite{thompson95,kostic96,geim98nat} and was considered as an
intrinsic property of any finite-size superconductor. Using the Ginzburg-Landau equations to analyze the flux capture and its compression, the
paramagnetic Meissner effect was predicted for mesoscopic conventional superconductors\cite{vvm97pme}, which was later on confirmed
experimentally\cite{geim98nat}. Besides flux compression\cite{vvm97pme,zharkov01}, the hysteresis transitions\cite{palacios98rapid,palacios00}, and the
metastability of the vortex configuration due to the influence of the sample surface\cite{deo99prb} were also considered.

In this paper, the magnetization of a superconducting mesoscopic square and triangle will be investigated by Hall magnetometry. The stability of the
vortex patterns imposed by the sample geometry  and the vortex expulsion and penetration will be discussed.

\section{Sample characteristics}

\begin{figure}[htb!]
\centering
\includegraphics*[width=4cm,clip=]{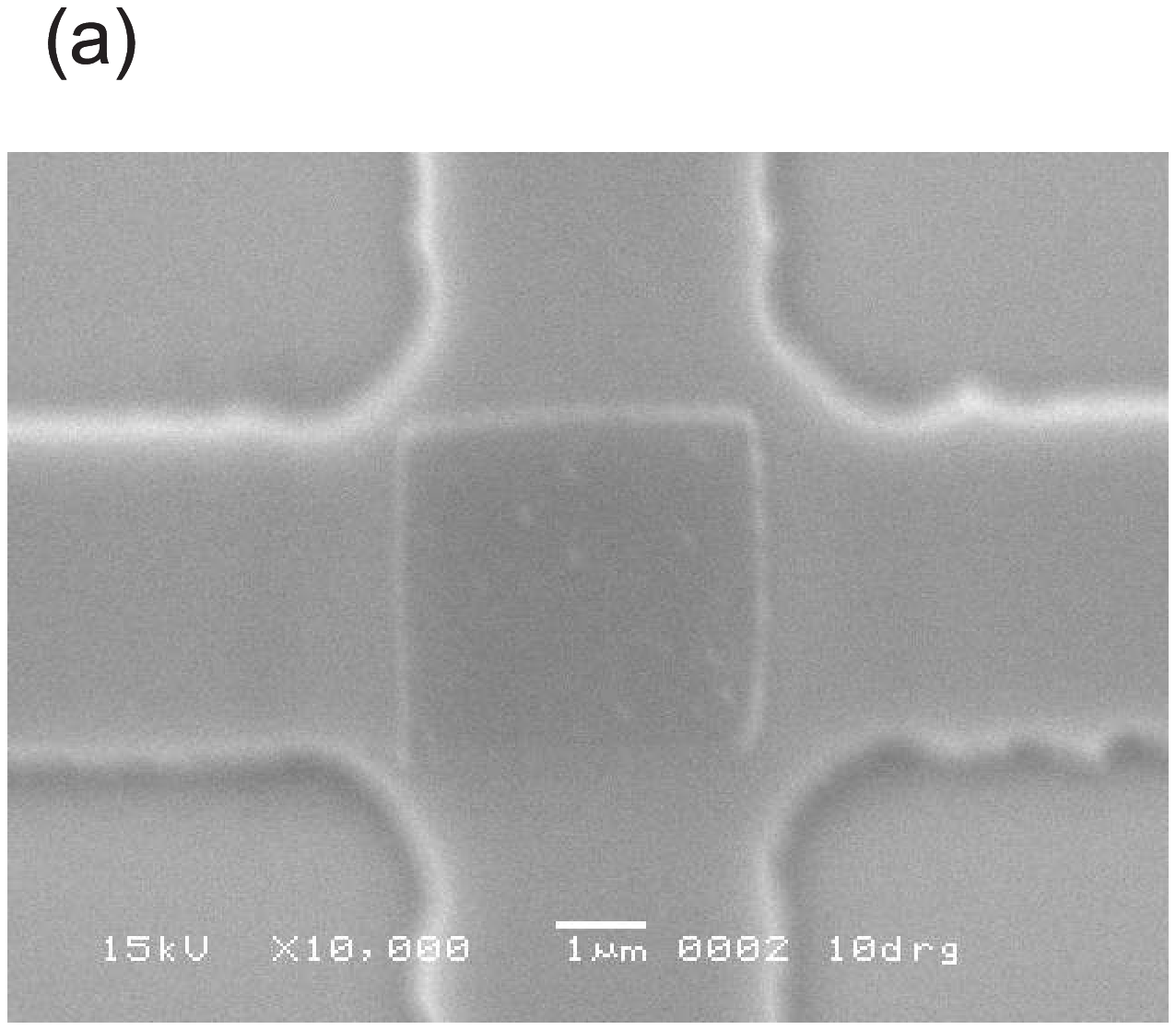}
\includegraphics*[width=4cm,clip=]{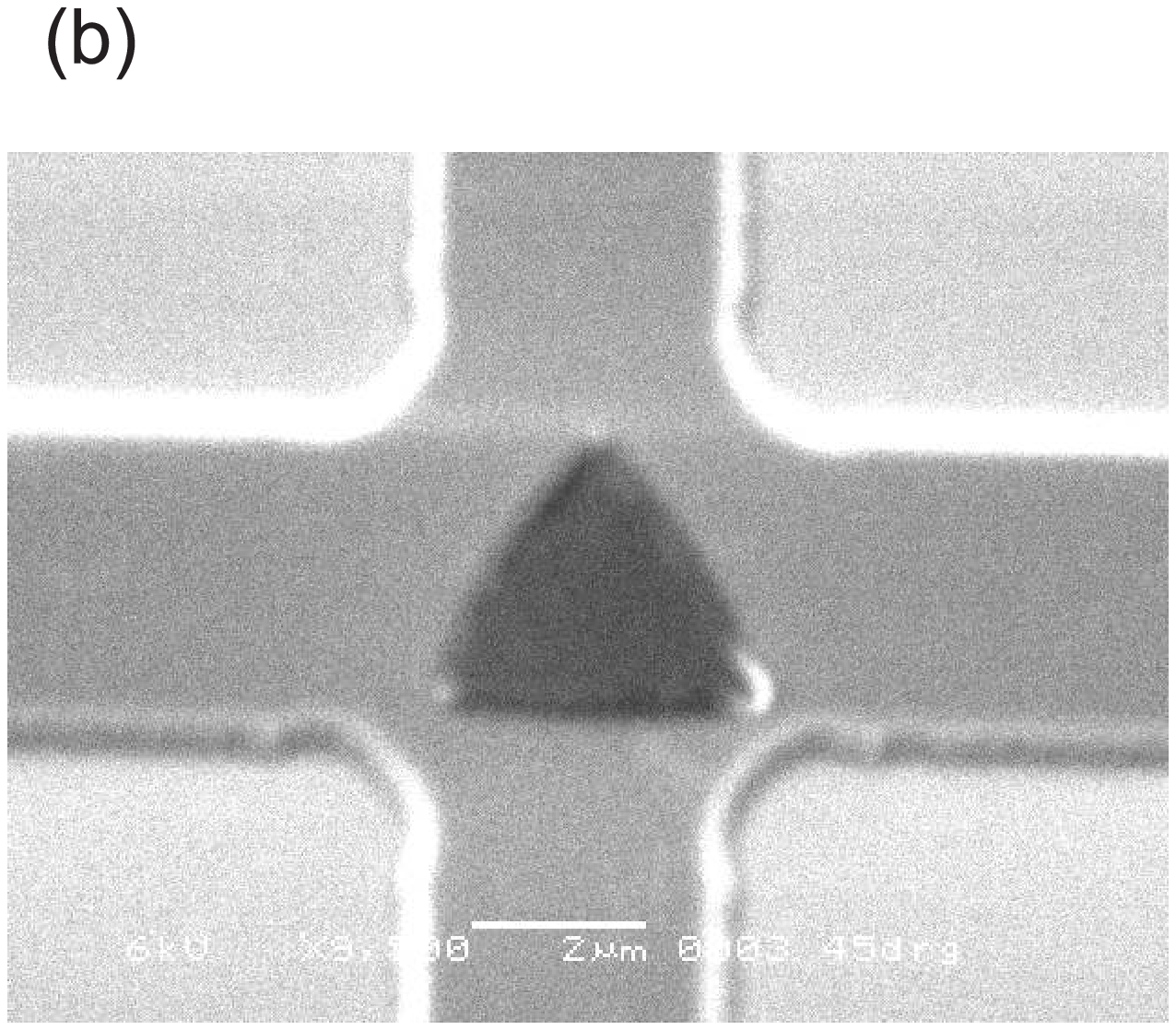}
\caption{SEM micrograph of the two Hall magnetometers. A superconducting Al square (a) and triangle (b) are deposited on the sensitive area of the Hall
probes.} \label{Fig:SEMSqTri}
\end{figure}

The micron Hall sensors  shown in Fig.~\ref{Fig:SEMSqTri} were fabricated from GaAs/AlGaAs heterostructures with the two-dimensional electron gas at a
distance $l$=70~nm below the surface. The active area is patterned by optical lithography. After etching, the Hall crosses were 3.8~$\mu$m wide giving a
sensing area of $3.7\times3.7~\mu$m$^2$ due to depletion of carriers at the edges. At the low temperatures of our measurements, the carrier mobility is
650000~cm$^2$/Vs and the 2D carrier concentration is $2.5 \times 10^{11}$~cm$^{-2}$, yielding a sensitivity of 0.25~$\mu$V per $\mu$AG. Our measurement
noise corresponds to 10 mG. The two Al structures were evaporated in separated runs, leading to slightly different samples parameters. The thickness of
the square was found from X-ray diffraction and from AFM to be $\tau=62$~nm. The thickness of the triangle was $\tau=38$~nm. The areas were
$S=14.5~\mu$m$^2$ for the square and $S=7.8~\mu$m$^2$ for the triangle. From macroscopic samples evaporated in the same runs, the coherence lengths of
$\xi(0)$=160~nm and $\xi(0)$=120~nm were determined for the square and the triangle, respectively.

Different values of the ac current applied to the Hall probes were used varying from 7.5 to 40 $\mu$A. Tuning the current and increasing the carrier
density by illuminating the Hall probes at low temperatures with an infra-red LED enables us to increase the sensitivity of the sensors.

\section{Experimental results}

\begin{figure}[htb!]
\centering
\includegraphics*[width=6cm,clip=]{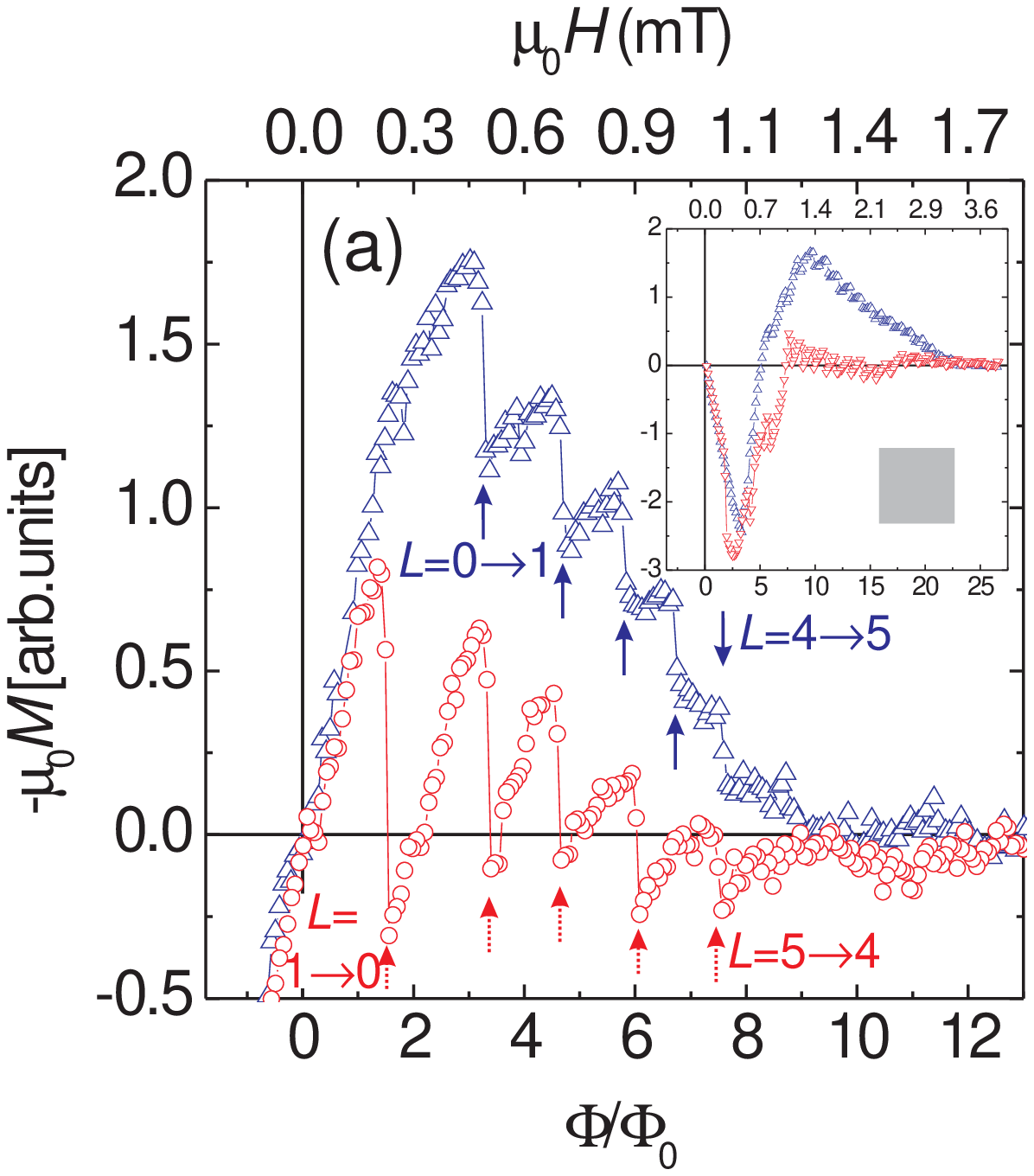}
\includegraphics*[width=5.9cm,clip=]{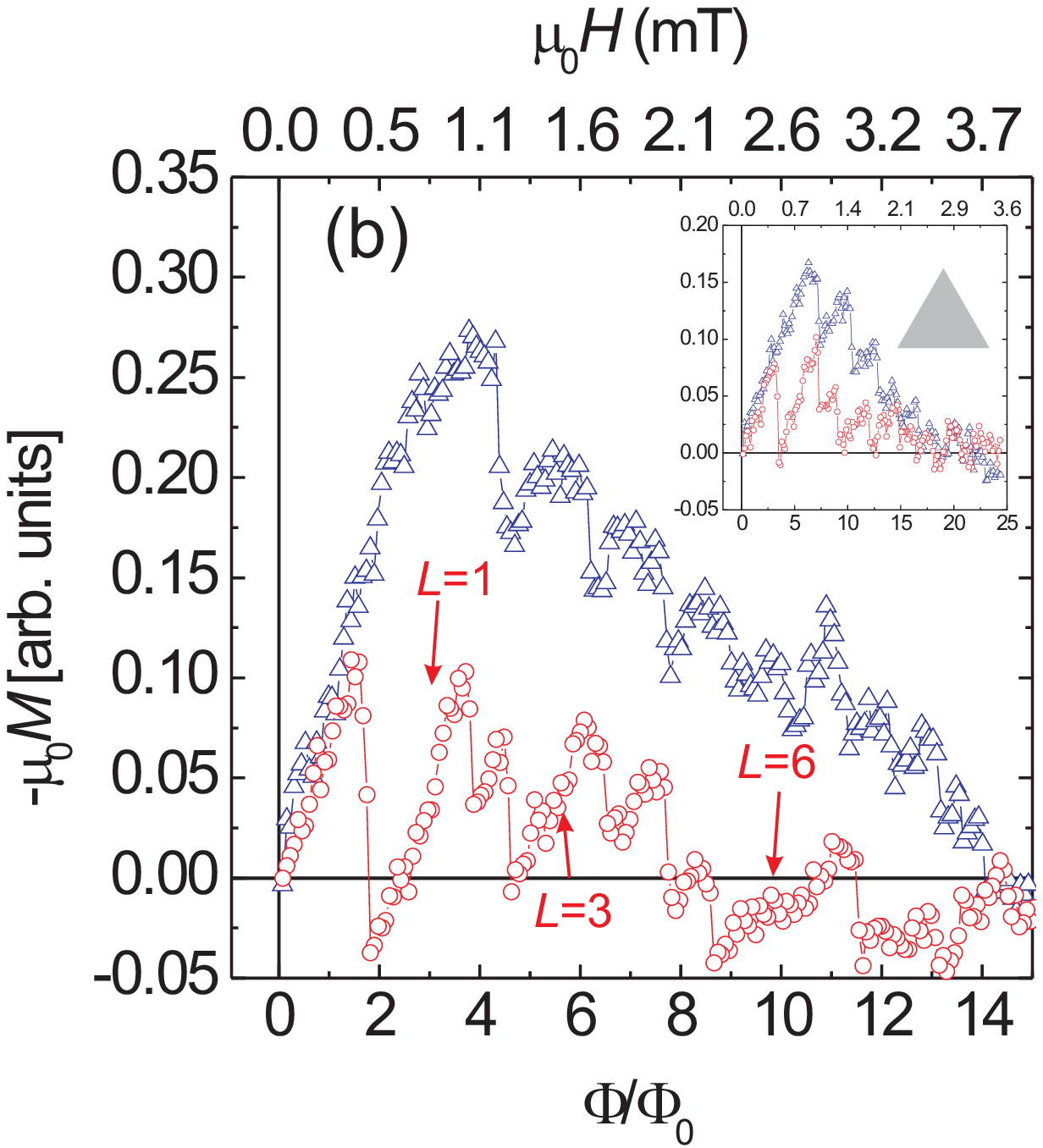}

\caption{(Color online) (a) Experimental magnetization curves of the Al square for the increasing (open squares) and the decreasing (open circles)
magnetic field. The measurements are performed at $T$=1.1~K (inset $T$=1.2~K). The full and dotted arrows show the magnetic fields used for the
determination of the $H-T$ diagram in Fig~\ref{Fig:Diagram}. (b) Experimental magnetization curves of the Al triangle at $T$=1.12~K (inset $T$=1.18~K).
$\Phi=\mu_0 H S$ denotes the flux through the sample area S and $\Phi_0$ is the flux quantum.} \label{Fig:MB}
\end{figure}

Fig.~\ref{Fig:MB} (a) shows the magnetization curve for the mesoscopic square at $T$=1.2 K ($T$=1.2 K in the inset). At the lowest temperature, the curve
exhibits a peculiar behavior at low magnetic fields with a strong paramagnetic signal. This paramagnetic dip has also been observed with empty Hall
probes. The origin of the signal was found to be from the wire bonds used to connect the current and voltage leads  from the Hall probe to the contacts
on the sample holder. These wires are made up from Al so that they can expel the magnetic field and increase the local field on top of the Hall sensor,
giving rise to a paramagnetic response. The four bonds are at a distance of approximately 100~$\mu$m from the sensing area of the Hall probe so that when
the Al wires are in the superconducting state, a large contribution to the Hall signal can be induced by the bonds. This problem can be solved by using
Au wires to replace the Al wires. Due to this extra contribution, the exact shape of the magnetization curve could not be obtained. Removing the signal
of a reference Hall sensor could not solve this problem. \emph{The paramagnetic signal will negligibly affect the position of the jumps} corresponding to
the penetration and expulsion of the vortices since the dip has an amplitude corresponding to $\sim$~0.03~mT.

More than 15 jumps in the increasing as well as in the decreasing branch of the $M(H)$ curve were seen corresponding to a maximal vorticity $L>15$ at
$T$=1.1~K for the square. Above 1.2~mT, where the critical field of the wire is exceeded, only the magnetization of the square is measured. The slope of
the parts with the same vorticity in the increasing branch are different than in the decreasing branch. This is in good agreement with the calculated
magnetization curve of Baelus {\it et al.\/}\cite{baelus02}, who determined the magnetization of superconducting disks, squares and triangles in the
framework of the nonlinear Ginzburg-Landau theory. They found in their calculations even a negative slope, what is not observed in our measurements for
the square. The origin of this discrepancy could be due to the different parameters used in the calculation or due to the presence of defects at the
boundary of the square. It is well known that the Bean-Livingston barrier\cite{bean64} is responsible for the hysteretic behavior in the magnetization of
superconductors\cite{cody66,joseph64,deblois64,deo99prb}. Geim {\it et al.\/}\cite{geim00nat} have shown experimentally, by introducing artificial
defects in a mesoscopic disk and by measuring the magnetization by means of a Hall sensor, that defects strongly decrease the Bean-Livingston barrier,
leading to a faster vortex penetration and expulsion. Small inhomogeneities in our sample would lead to a lower value of the penetration field so that
the calculated negative slope for a perfect square would not be observed in our measurements.

Above 1.2~K, no extra signal from the Al bonds was measured. This indicates that the critical temperature of the wires is comparable to that of bulk Al
(1.196~K). At this temperature [see Fig.~\ref{Fig:MB} (a)], a vorticity up to $L$=5 was observed in the square. The exact shape of the magnetization
curve was not perfectly reproduced. The positions of the transitions from $L$ to $L+1$ were however well reproduced since \emph{the accuracy on the
determination of the position of the jumps is not affected by the noise of the Hall probe but only by the accuracy of the applied field.} The transition
from $L=0$ to $L=1$ in the increasing branch and from $L=2$ to $L=1$ in the decreasing branch occur at approximately the same magnetic field. We should
then expect a continuous curve for the state with vorticity $L=1$. This was observed in some measurements but not always reproduced. The somewhat
contradictory observations results, most probably, from the not perfectly linear response of the Hall voltage as a function of the magnetic field.
Actually, above $T_c$, a weak hysteretic behavior of the Hall voltage was sometimes measured. Removing this effect from the data is not easy since the
hysteretic signal was not always reproduced.

\emph{A paramagnetic signal was observed for the decreasing branch of the magnetization.} It is difficult to extract directly from the experiment whether
the paramagnetic behavior is caused by the specific detector geometry or whether it has a physical origin. Since the square is filling the largest part
of the sensing area, we believe that the Hall sensor itself is not responsible for the paramagnetic response\cite{deo99prb}.

The magnetization curves of the triangle for increasing and decreasing fields are shown in Fig.~\ref{Fig:MB} (b) for $T$=1.12~K (inset $T$=1.18~K). The
triangle has an area approximately two times smaller than the square, so that the maximal vorticity for the lowest temperature observed in the triangle
is $L$=8. Contrary to the case of the square, a negative slope is found in the increasing branch of the curve, indicating a strong penetration of the
magnetic field inside the triangle. This is in agreement with the calculations reported in Ref.~\onlinecite{baelus02}. At higher temperatures, the
curvature at the end of each line is less pronounced and evolves to an almost linear behavior at $T$=1.18~K. Due to the curvature, the jumps occurring at
the increasing magnetic field are poorly resolved. Since the jumps are easily resolved in the decreasing branch, a possible way to detect the transition
for the increasing magnetic field would be to sweep constantly the field up and down to zero and increase each time slightly the maximal magnetic field.
In this way, the vorticity at the maximal field could be estimated from the decreasing part of the curves. For the two temperatures presented here,
\emph{a paramagnetic Meissner effect has been observed} for some vorticities. While the vortices enter the sample in an almost periodic way, the
decreasing branches of the magnetization curves show all a non-periodic behavior. The lines with vorticity $L$=1 and 3 are much longer than the others.
Also the line with $L$=6 at $T$=1.12~K is very long. It indicates a very stable configuration at these vorticities, which can be a consequence of the
geometry that forces the vortices to keep the symmetry of the sample.

%

\begin{figure*}[htb!]
\centering
\includegraphics*[width=5.9cm,clip=]{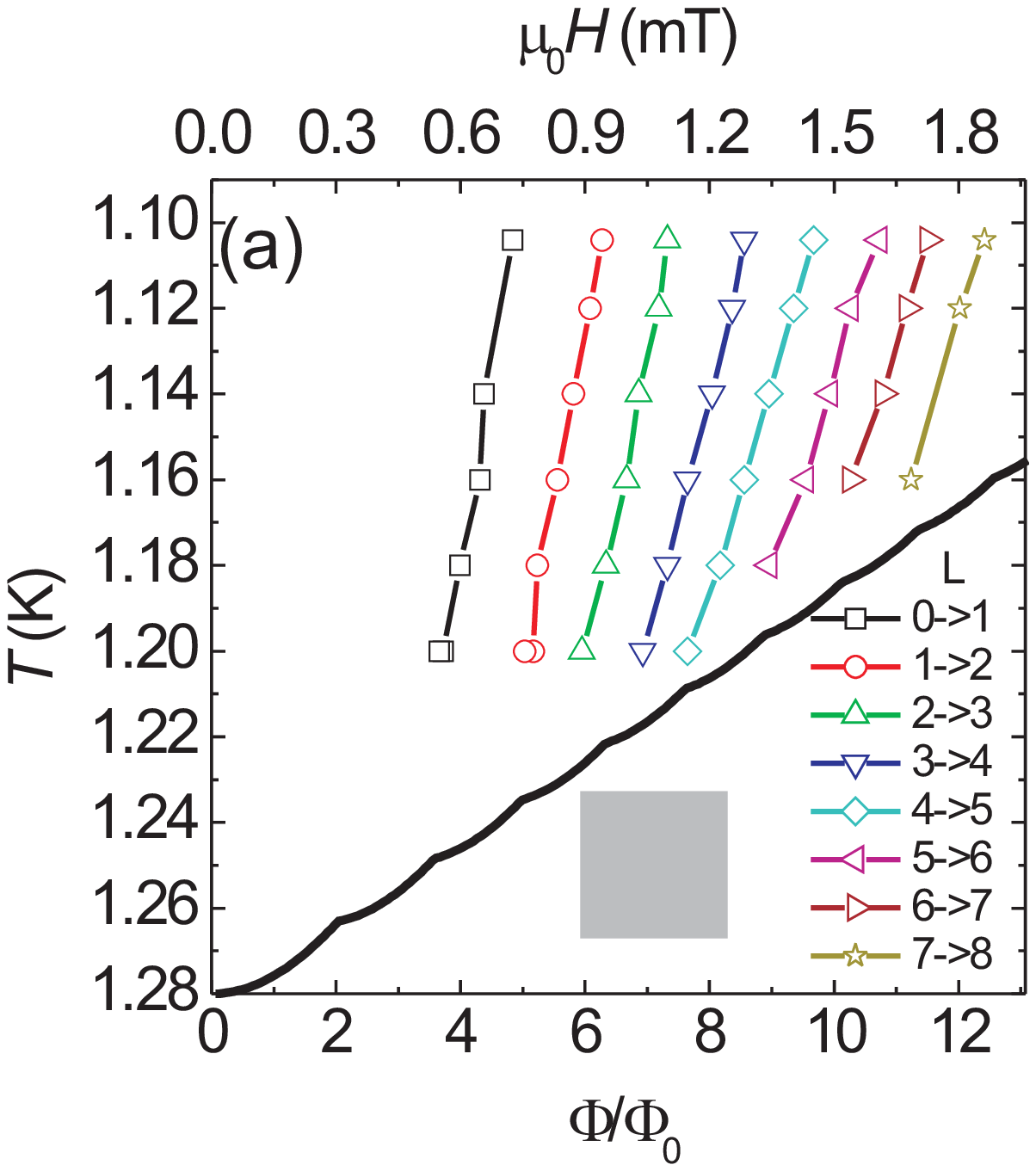}
\includegraphics*[width=5.9cm,clip=]{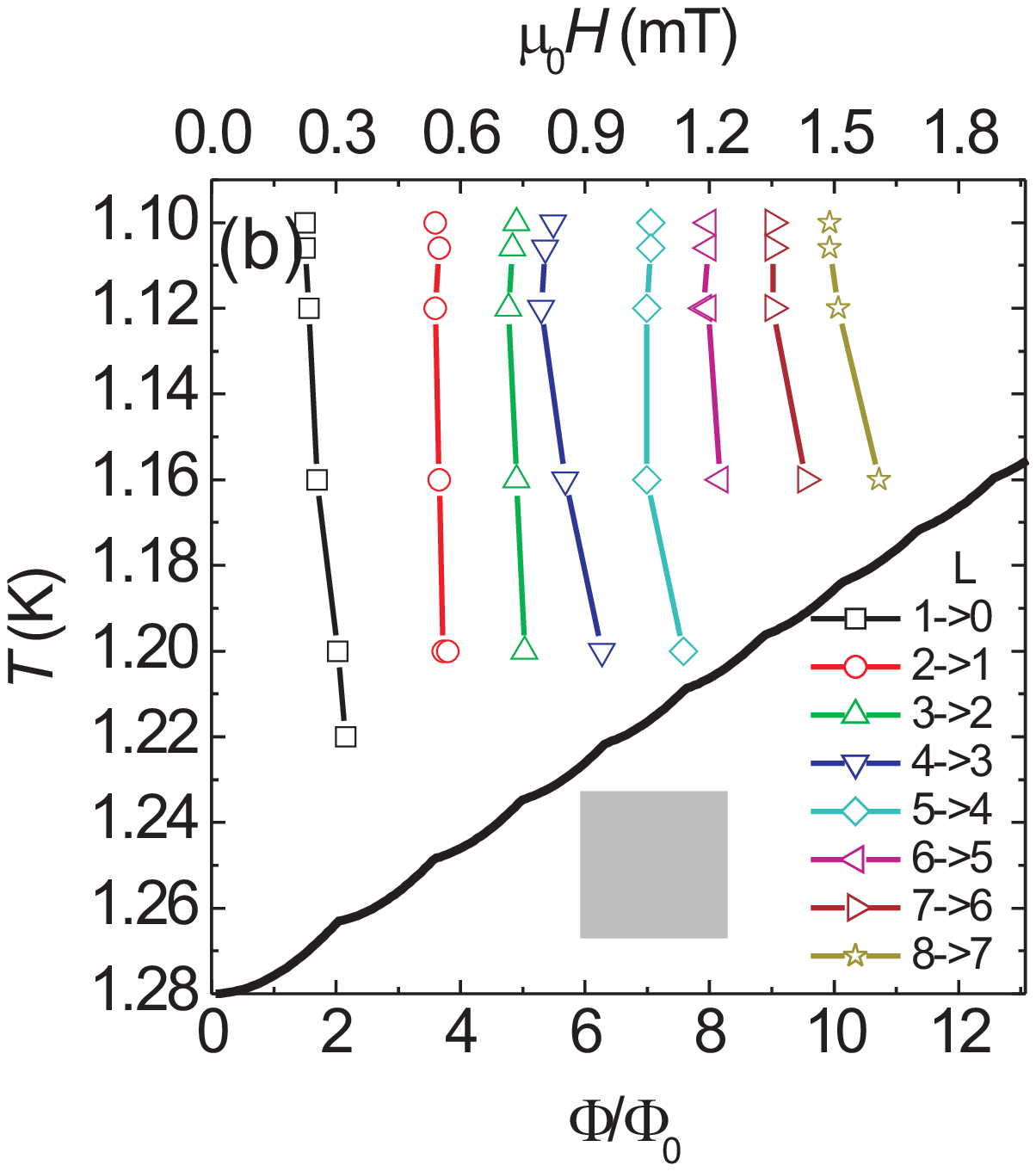}
\includegraphics*[width=5.9cm,clip=]{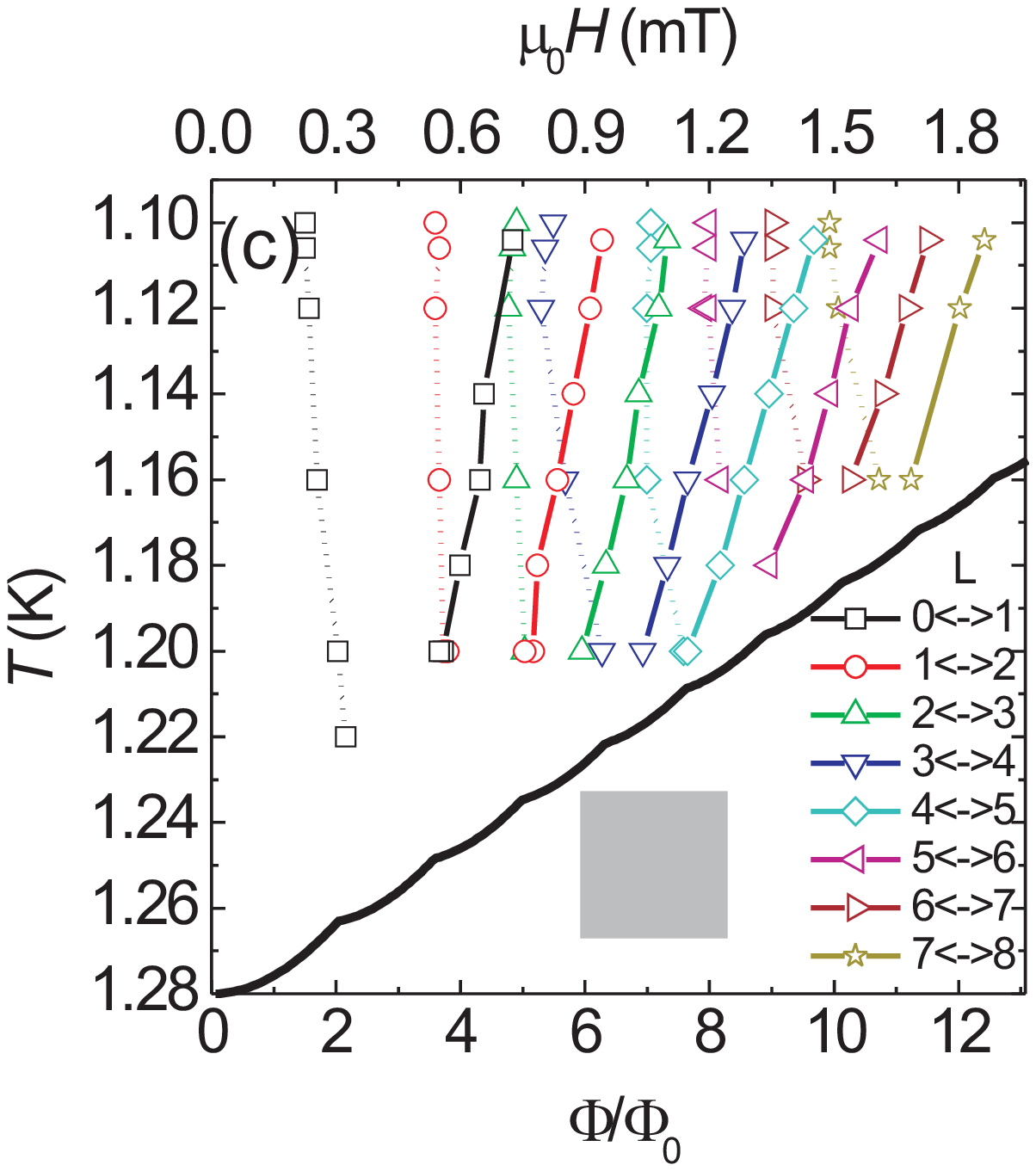}
\includegraphics*[width=5.9cm,clip=]{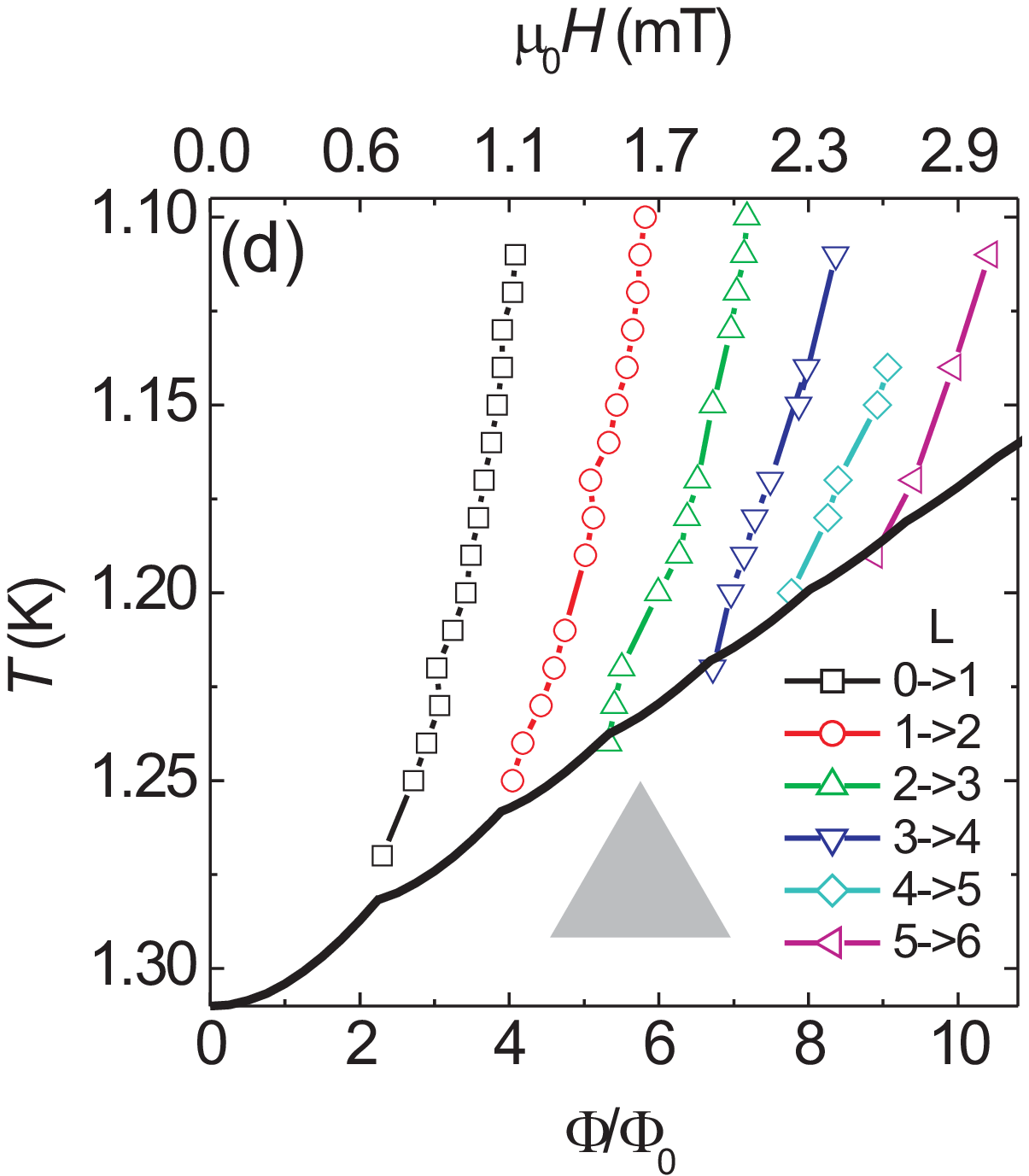}
\includegraphics*[width=5.9cm,clip=]{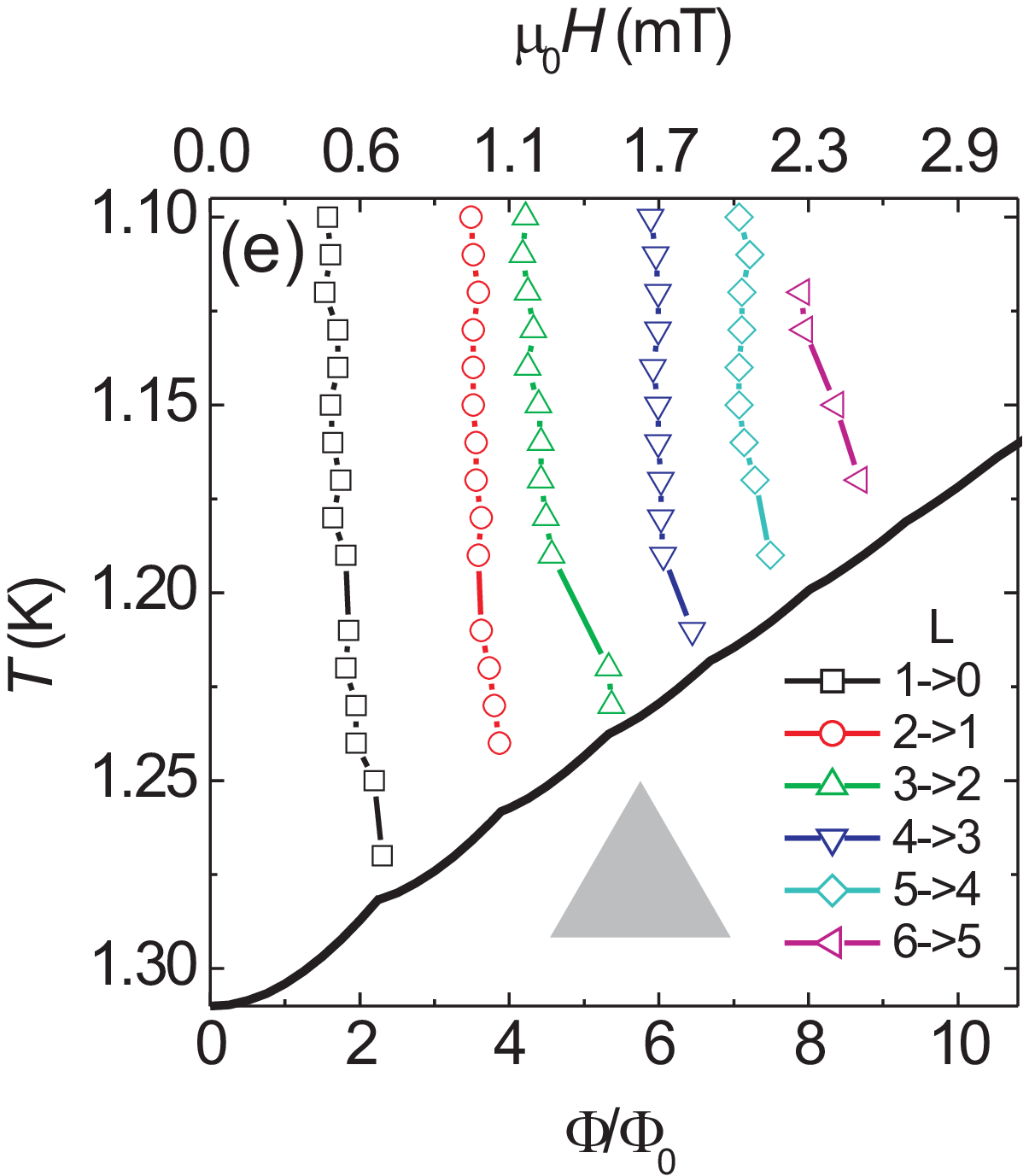}
\includegraphics*[width=5.9cm,clip=]{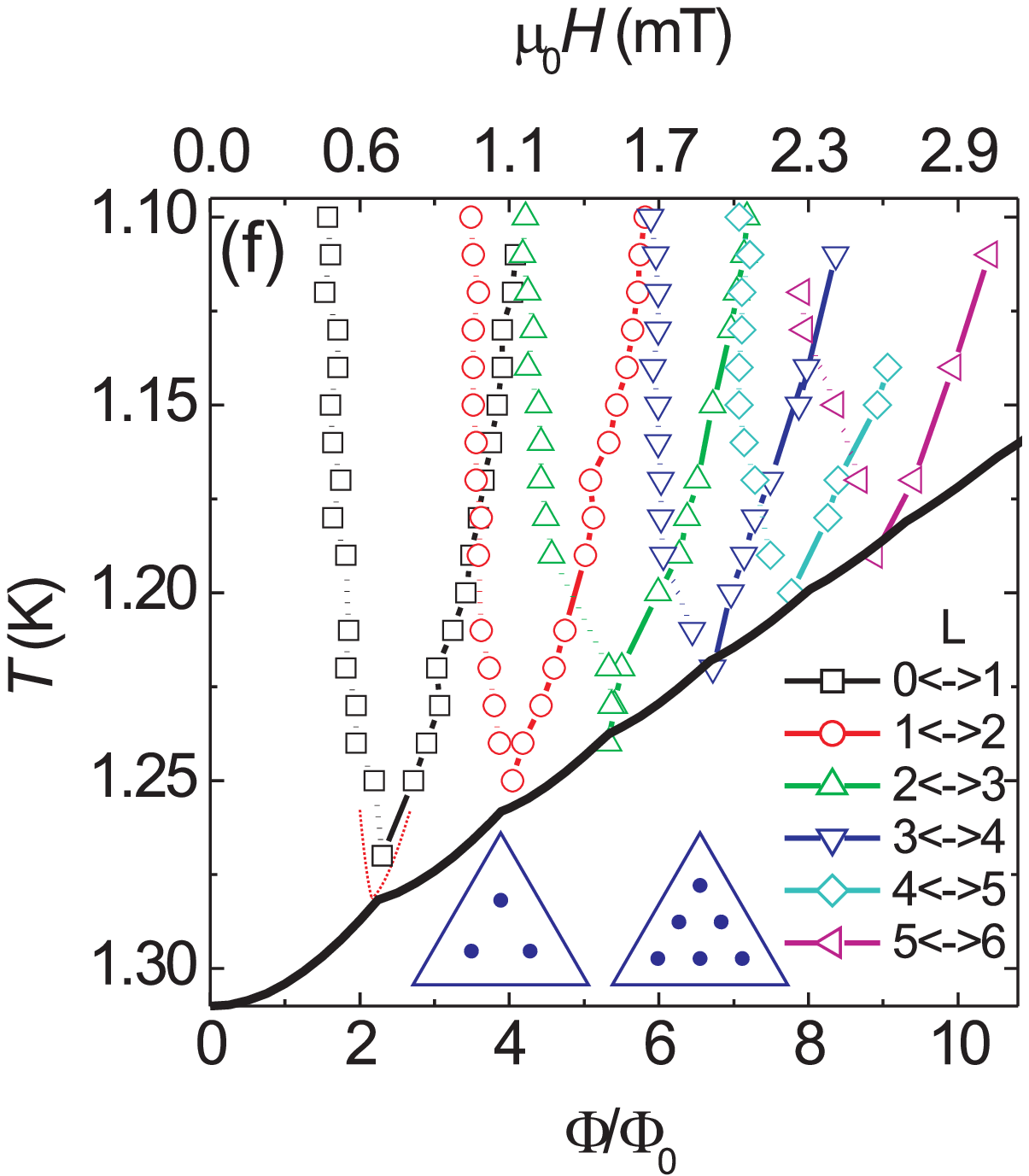}
\caption{(Color online) $H-T$ diagram of the square (a) for the vortex entry, (b) for the expulsion and (c) for both the entry and expulsion. At the thin
solid line, a vortex penetrates the sample and is expelled at the dotted line. The transitions are shown from $L=0\leftrightarrow 1$ (open squares) to
$L=7\leftrightarrow 8$ (open stars). The lowest solid cusp-like line (thick line) represents the theoretical phase boundary of a square. The area
$S$=13.5~$\mu$m$^2$, the coherence length $\xi(0)$=~180~nm and the critical temperature $T_c$=1.28~K were therefore used. (d-f) idem but for the
triangle. The area $S$=7.2~$\mu$m$^2$, the coherence length $\xi(0)$=~170~nm and the critical temperature $T_c$=1.31~K were used for the calculation of
the theoretical phase boundary of the triangle. The thin line in (f) shows the theoretical calculated transition\cite{baelus02} between the states
$L=0\leftrightarrow 1$ close to the $T_c(H)$ phase boundary and the inset shows a possible vortex configuration at $L=3$ and $L=6$.} \label{Fig:Diagram}
\end{figure*}


The magnetization has been measured at different temperatures. By monitoring the transitions between states with different vorticities, a $H-T$ diagram
has been reconstructed (see Fig.~\ref{Fig:Diagram}) showing the penetration (a,d) and expulsion (b,e) of a vortex. For the vortex penetration in the
square [Fig.~\ref{Fig:Diagram} (a)], the lines corresponding to different vorticity are all more or less parallel and equidistant to each other. It can
be seen that the expulsion of the last vortex from the sample takes place at a much lower field value, meaning that the state with $L$=1 is very stable.
The curve corresponding to the transition from $L$=4 to $L$=3 [$\triangledown$ in Fig.~\ref{Fig:Diagram} (b)] has a strongly different shape than the
others. At low temperatures, the expulsion of a vortex is delayed for this transition, while it is not so pronounced at higher temperatures. The
configuration with four vortices is a stable configuration in the square as to be expected. Indeed, at $L$=4, the vortices will be configured so as to
preserve the $C_4$ symmetry imposed by the boundary conditions\cite{chibotaru00,bonca01,baelus02}. It seems that the penetration occurs in an almost
periodic way, only weakly dependent on the vorticity, while the expulsion is delayed for $L$=4 and $L$=1. \emph{This indicates that the penetration
mechanism is less dependent on the vortex configuration inside the sample than the expulsion.} A possible qualitative explanation for these two different
behaviors is that for the expulsion, vortices, which are in the corners of the square in the case of a stable configuration, will first have to move to
the middle of the side and then cross the barrier. The motion from the corners to the side will cost energy since the exiting vortex is repelled by the
other vortices sitting on the remaining corners. Another possibility is that they cross the barrier around the corners without disturbing too much the
vortex configuration, but the barrier is higher there. For the penetration, the vortex configuration inside the sample will only be affected strongly
once the new vortex has crossed the barrier.

The $H-T$ diagram  showing the vortex penetration and expulsion in the triangle is presented in Fig.~\ref{Fig:Diagram} (d-e). Similar to the square, the
lines corresponding to the penetration of a new vortex into the triangle run parallel and are almost equidistant. This indicates that the penetration
mechanism is almost not affected by the vortex configuration inside the triangle. The expulsion lines, contrary to the penetration lines, are strongly
aperiodic. The expulsion of a vortex when the vorticity is $L$=1 and $L$=3 is strongly delayed. It is interesting to note that the slope of the line
corresponding to the transition from $L=6$ to $L=5$ is different than the others. It was also seen in Fig.~\ref{Fig:MB} (b) that the line with vorticity
$L$=6 is very long at $T$=1.12~K. A possible reason for this observation is that at low temperature, a vortex molecule is formed where the six singly
quantized vortices are forming a triangular lattice [see inset in Fig~\ref{Fig:Diagram} (f)]. This configuration is expected to be very stable since it
keeps the symmetry imposed by the geometry of the sample and it forms an Abrikosov triangular lattice, which gives the lowest energy due to the maximal
spreading of the vortices. For higher temperatures, the giant vortex state will be more favorable. The vorticity will probably affect less strongly the
expulsion mechanism when in the giant vortex state. This would result in an expulsion at higher magnetic field value for higher temperatures.

It is worth emphasizing that the position where the entry and exit lines join corresponds to the position of the cusps in the theoretical phase boundary
[Fig~\ref{Fig:Diagram} (f)]. This is expected since at the $T_c(H)$ line only one vorticity can exist at a given magnetic field and the transition
between two vorticities will happen exactly at the position of the cusps as well for increasing as for decreasing field. These results are in good
agreement with the calculation of the penetration and expulsion fields calculated for $L=0\leftrightarrow 1$ in Ref~\cite{baelus02} [thin line in
Fig.~\ref{Fig:Diagram} (f)].

\begin{figure}[hbt!]
\centering
\includegraphics*[width=6cm,clip=]{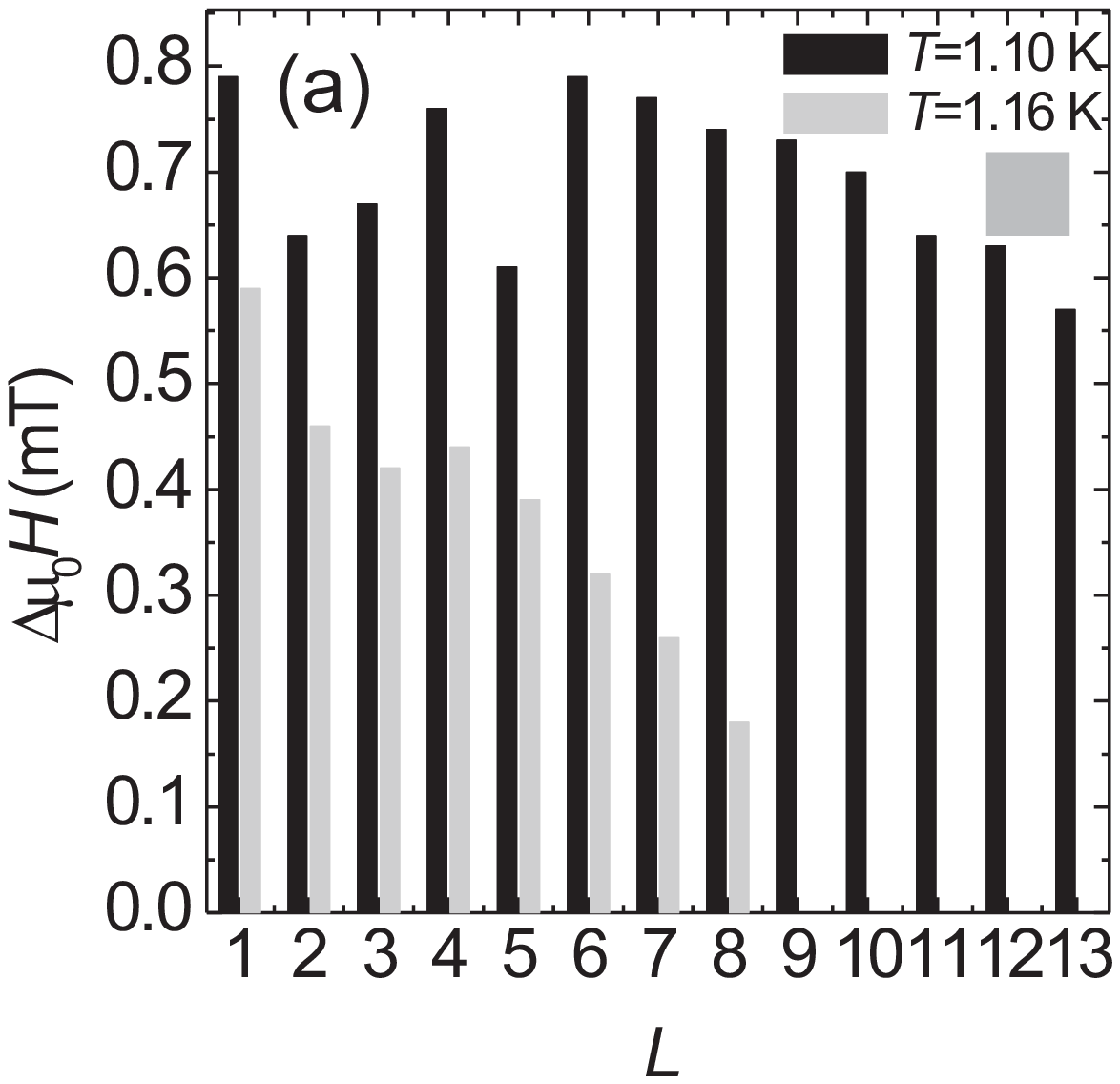}
\includegraphics*[width=6cm,clip=]{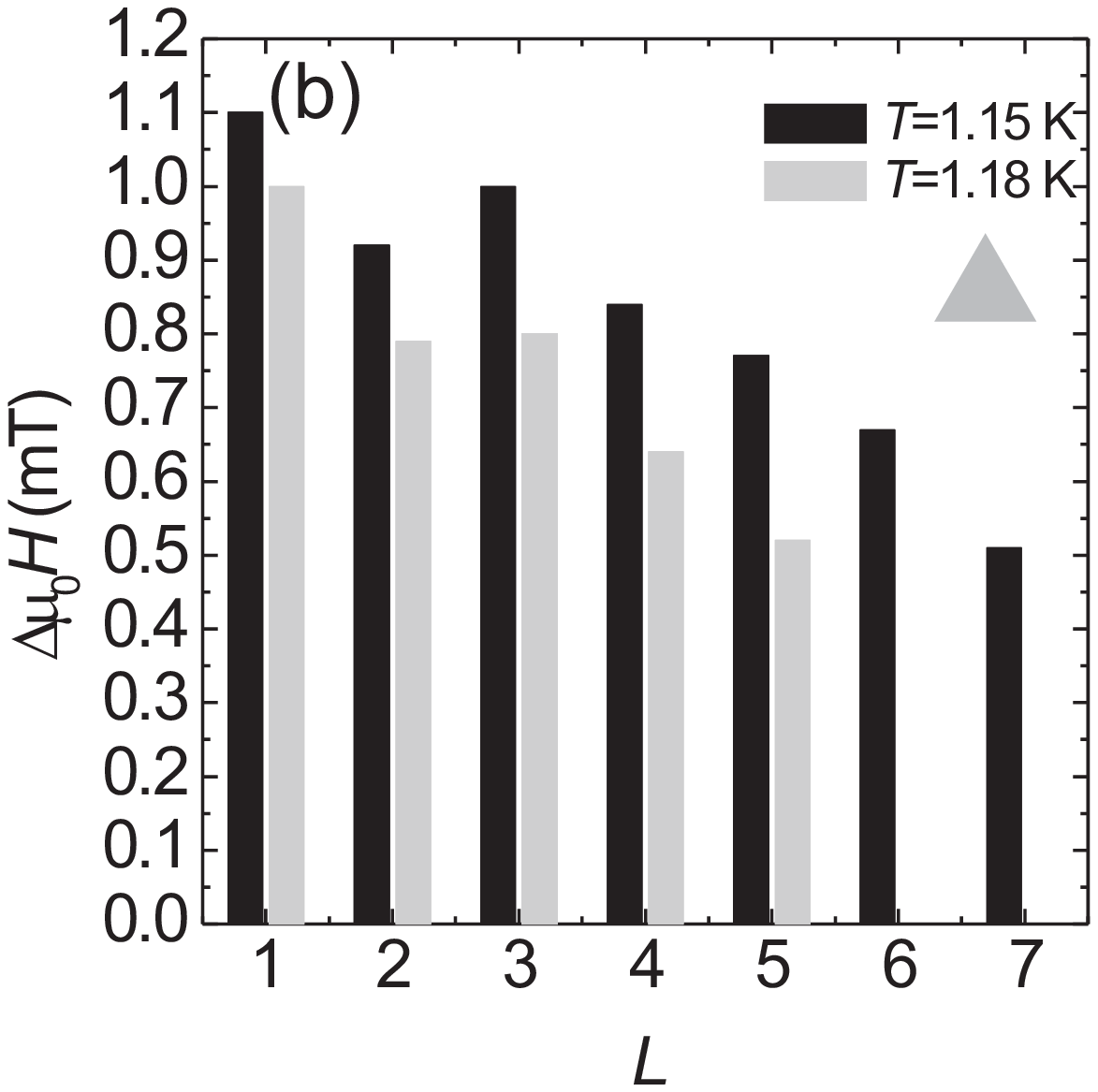}
\caption{The magnetic field range $\Delta\mu_0H=\mu_0(H_{L\rightarrow L+1}-H_{L\rightarrow  L-1})$ over which the state with vorticity $L$ is stable (a)
for the square at $T$=1.1~K (black bars) and $T$=1.16~K (grey bars) and (b) for the triangle at $T$=1.15~K (black bars) and $T$=1.18~K (grey bars).}
\label{Fig:DeltaH}
\end{figure}

In order to study  different vortex patterns more carefully, the stability region $\Delta H=H_{L\rightarrow L+1}-H_{L\rightarrow L-1}$ where a state with
constant vorticity has been measured is shown in Fig.~\ref{Fig:DeltaH} for different temperatures. The square [Fig.~\ref{Fig:DeltaH} (a)] displays a peak
structure as a function of the vorticity for the two presented temperatures. At high vorticity, the stability region decreases almost uniformly. A stable
configuration is found at $L$=4, as a consequence of the fact that the vortex lattice tries to keep the geometry imposed by the boundary. What is
interesting to notice is that the state with $L$=6 remains stable over a longer range than the state with $L$=4 at the low temperature. Above this value,
no peak has been seen, only small steps at $L$=10 and $L$=12. It was shown using the linearized GL equation\cite{chibotaru00} that the vortex
configuration at $L$=6 consists of a giant vortex with vorticity of 2 in the middle and with 4 singly quantized vortices located on the diagonals. This
configuration is only valid very closed to the phase boundary, where the solution has to follow the $C_4$ symmetry of the square. The vortex pattern deep
in the superconducting state can be totally different. There, the linearized GL equation is not valid anymore and the nonlinear term has to be taken into
account in order to solve the problem. Symmetry-breaking solutions can exist there\cite{mertelj03,TeniersPhD}. A possible explanation why the regions
with constant vorticity are continuously decreasing above $L$=6 is that only giant vortices are formed for high vorticity. No preferred value for the
vorticity has to be expected in this case. Baelus {\it et al.\/}\cite{baelus02} found indeed vortex molecules up to $L$=6. For $L$=2, 3 and 6, a
transition from the state with separated vortices to the giant vortex state was found when increasing the magnetic field. Above $L$=6 only giant vortex
states were found. It is important to mention that their calculations were performed with different parameters than the ones suitable for our experiment,
so that no quantitative comparison can be performed.

In Fig.~\ref{Fig:DeltaH} (b) the magnetic field range $\Delta H$ over which a state with vorticity $L$ is presented for $T$=1.15~K and $T$=1.18~K for the
triangle. Unfortunately, the same analysis could not be performed at lower temperatures, where a higher vorticity is obtained, due to the difficulty to
resolve the penetration of new vortices in the triangle at lower temperatures. There, the curvature of the magnetization is strongly pronounced so that
the height of the jumps in the increasing branch is severely reduced. For the two temperatures presented in Fig.~\ref{Fig:DeltaH} (b), a peak is observed
for $L$=1 and $L$=3, as expected from previous discussions. Above $L$=3, the stability region $\Delta H$ decreases monotonously. No other stable
configurations were observed for higher fields.

\section{Conclusion}
The magnetization of a mesoscopic Al square and triangle was measured. The stability of different vortex patterns was analyzed. The magnetization vs
field curves were compared with previous theoretical calculations for a square and a triangle. A good qualitative agreement was obtained. For some
vorticities a paramagnetic signal was seen. We found that the penetration of a new vortex in the sample is almost not affected by the vortex
configuration inside the superconductor, while the vortex expulsion is strongly controlled by the stability of the vortex pattern. A stable configuration
with $L$=4 was found for the square and with $L$=3 for the triangle. At higher vorticity and at higher temperature, a continuous decrease of the magnetic
field range $\Delta H$, over which a state with constant vorticity exist, was seen. We attribute this effect to the formation of a giant vortex state.

\begin{acknowledgments}
This work has been supported by the Belgian IUAP, the Flemish FWO, the Research Fund K.U.Leuven GOA/2004/02 programmes and by the ESF programme VORTEX.
M.M. is a postdoctoral fellow of IWT-Vlaanderen. The authors would like to thank O. Popova for the X-rays measurements.
\end{acknowledgments}

\end{document}